\documentclass[3p,twocolumn]{elsarticle}

\usepackage{amssymb}
\usepackage{amsmath}
\usepackage{lineno}
\journal{Chaos Solitons and Fractals}

\begin{document}
\begin{frontmatter}

\title{Social dilemmas in off-lattice populations}

\author[label1]{B.F. de Oliveira} 
\author[label2]{A. Szolnoki}

\address[label1]{Departamento de F\'\i sica, Universidade Estadual de Maring\'a, 87020-900 Maring\'a, PR, Brazil}
\address[label2]{Institute of Technical Physics and Materials Science, Centre for Energy Research, P.O. Box 49, H-1525 Budapest, Hungary}

\begin{abstract}
Exploring the possible consequences of spatial reciprocity on the evolution of cooperation is an intensively studied research avenue. Related works assumed a certain interaction graph of competing players and studied how particular topologies may influence the dynamical behavior. In this paper we apply a numerically more demanding off-lattice population approach which could be potentially relevant especially in microbiological environments. As expected, results are conceptually similar to those which were obtained for lattice-type interaction graphs, but some spectacular differences can also be revealed. On one hand, in off-lattice populations spatial reciprocity may work more efficiently than for a lattice-based system. On the other hand, competing strategies may separate from each other in the continuous space concept, which gives a chance for cooperators to survive even at relatively high temptation values. Furthermore, the lack of strict neighborhood results in soft borders between competing patches which jeopardizes the long term stability of homogeneous domains. We survey the major social dilemma games based on pair interactions of players and reveal all analogies and differences compared to on-lattice simulations.
\end{abstract}

\end{frontmatter}

\section{INTRODUCTION}

In a multi-agent system the assumption when every member interacts with all others randomly can be handled analytically, hence it could always be a starting point to study the evolution of cooperation among self-interest players \cite{hofbauer_88,nowak_06,wang_sx_pla21,shao_yx_epl19,xu_zj_c19}. The absence of stable connections, however, is a highly simplified working hypothesis because in almost every real-life examples individuals have fixed, or at least temporarily stable neighbors \cite{newman_siamr03,eguiluz_ajs05}. This observation can be modeled by assuming an interaction graph where players have limited and stable partners, which fact determines their potential fitness and the dynamical process in the applied topology \cite{nowak_n92b,ahmed_epjb00,santos_prl05}. As expected, this modification may change the system behavior significantly which was confirmed by thousands of research papers in the last two decades \cite{szabo_pr07,amaral_rspa20,liu_xs_epl19,roca_plr09,yang_gl_pa19,perc_pr17,liu_rr_amc19,jiao_yh_csf20,amaral_pre20}.

Notably, the graph approach does not always reflect faithfully the interactions of individuals. For instance, in a microbial environment an off-lattice approach seems to be more appropriate modeling technique, where individuals still interact with a limited number of partners, but their actual distance, which may change continuously, determines the interaction strength \cite{jansen_mb05,kreft_mb05,garde_rsob20}. In the last years several experimental works have been published where an off-lattice model seems to be a more realistic assumption \cite{drescher_cb14,boyle_pcbi15,tekwa_prsb17}.

We must stress, however, that off-lattice simulations are more demanding technically and requires significantly larger numerical efforts comparing to on-lattice or in more general graph-based simulations. Therefore it is not surprising that previous works were restricted to the latter case exclusively \cite{quan_j_pa21,nagatani_c20,zhang_lm_pa21,he_qp_epl20,li_k_csf21,gao_sp_pre20,quan_j_jsm20,yang_r_epjb20}. In our present work we focus on off-lattice simulations and explore their specific characters. Importantly, we use the original social dilemma games to identify the similarities and potential differences between on- and off-lattice environments and only change the dynamical rules which can be applied for off-lattice simulations directly. We will show that the behavior of systems in off-lattice environment is conceptually similar to those observed for models of lattice-type interaction graphs. There are, however, some differences which warn us to treat the conclusions of lattice-based models with a special care when we want to adopt them directly to microbiological or related systems. 

In the following we specify the model and the microscopic rules in more detail. After we present its consequences for prisoner's dilemma game, which is extended to the related social dilemmas including snow-drift and stag-hunt games. Finally we conclude with the summary of the results and a discussion of their implications in the last section.

\section{MODELING DILEMMAS IN OFF-LATTICE ENVIRONMENT}
\label{sec:model}

As we emphasized, we consider the traditional social dilemma games where players interact with their partners and collect payoff elements from every specific pair interaction \cite{sigmund_10}. A player's state can be described as a cooperator or defector. When two cooperator players meet then both of them obtain a payoff value 1, while the meeting of defectors yields zero payoff value for each participant. The interaction of a cooperator and a defector players provides a value $T$ (temptation) for the latter and $S$ value (sucker's payoff) for the former strategy. In this way, the actual values of $T-S$ pair determine the character of the social dilemma. By keeping the traditional parametrization we use $T>1$ control parameter and $S=0$ fixed to describe the (weak) prisoner's dilemma game. Furthermore, a parameter $0 \le r \le 1$ when $T=1+r$ and $S=1-r$ serves as a control parameter to span the snow-drift game region of the $T-S$ parameter plane. Last, the stag-hunt game region is covered by the same $0\le r \le 1$ control parameter when $S=-r$ and $T=r$ \cite{perc_bs10}.

In an off-lattice environment $N$ players are distributed randomly on a square-shaped box of linear size $L = 1$. As usual for spatial populations, periodic boundary conditions are applied. For every $k$ player the horizontal $x_k$ and vertical $y_k$ coordinates are continuous variables. A player $k$ and a player $m$ interact if they are within the interaction range, namely their distance is less than $l_i$. Note that to calculate the proper distance we consider the mentioned periodic boundary conditions.

To introduce an evolutionary dynamics we assume that a player's strategy may change according to the broadly applied pairwise comparison imitation dynamics \cite{szabo_pre98}.
More precisely, a player $k$ will change its strategy to the opposite strategy represented by a neighboring player $m$ with a probability $w$ that depends on the difference of the $\Pi_k$ and $\Pi_m$ payoff values collected by the mentioned players: 
\begin{equation}
w=\frac{1}{1+\exp[(\Pi_k-\Pi_m)/K]} \,\,.
\nonumber
\end{equation}
In this Fermi-function-type formula parameter $K$ represents a noise level, where the zero limit makes the change deterministic for positive difference and forbidden in the reversed case, while large $K$ limit provides a random strategy change independently of the actual payoff values of players.

It is important to stress that in an off-lattice simulation we cannot follow precisely the usual way of strategy adoption applied in graph-based simulations. In particular, we cannot simply replace the strategy of the target player with the new one and leave its position unchanged because this would make the evolutionary outcome highly sensitive on the initial spatial distribution of players. Instead, we remove the player who wants to change its strategy and add a newborn player with the new strategy somewhere randomly within an $l_b$ distance of the model player. This modification is similar to the so-called death-birth dynamics, but still keeps the essence of pairwise comparison and makes possible the match with graph-based simulations \cite{ohtsuki_jtb06}.

In the default case $l_b$ is chosen to be equal to $l_i$, but we also discuss smaller and larger ranges when adding a newborn player to replace the old one. As we will show the former option has no, while the latter change does have relevant consequence on the results. In general, it is important to emphasize that the evolutionary outcome could be significantly different even if we use the same parameter values of the model. For instance the system may be trapped into a homogeneous full defector state, but  we may also observe a coexistence phase by using another seed value of our random number generator. Therefore to obtain a reliable observation about the system behavior it is vital to average the results of individual runs. At every parameter values we distributed the competing strategies and the positions of $N$ players randomly and monitored how the fractions of strategies change in time. After we repeated every simulation $1000$ times to obtain the requested accuracy of averaged value.

\section{RESULTS}
\label{sec:result}

\begin{figure}
\centering
\includegraphics[width=7.8cm]{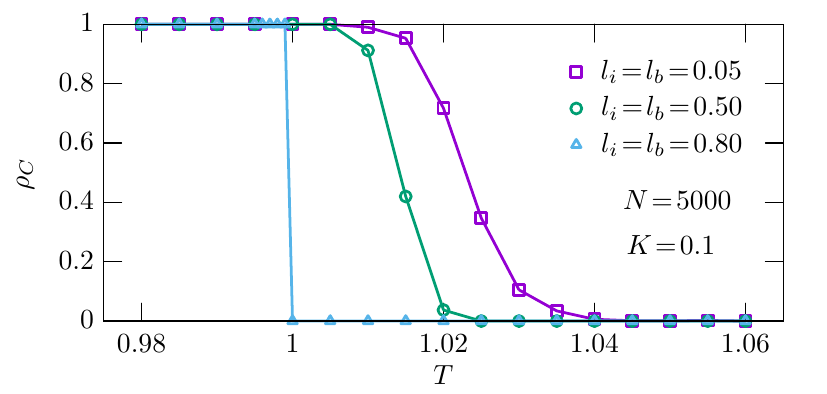}\\
\caption{Cooperation level in dependence of temptation $T$ for prisoner's dilemma. The applied characteristic length values are indicated for each curves. If $l_i=l_b \ge 0.8$ then we practically get back the behavior of a well-mixed system. Curves are just to guide to the eye.}\label{to-mix}
\end{figure}

First we summarize our observations obtained for prisoner's dilemma game and after we briefly outline the results of other dilemma games. But before jumping deep into the prisoner's dilemma case we first present the results obtained for different lengths of the interaction range. This helps us to get impressions and evaluate properly the characteristic length scale of the $l_i$ parameter. Figure~\ref{to-mix} shows the cooperation level in dependence temptation value for different characteristic length scales. Here we used the default model where the interaction ($l_i$) and reproduction ($l_b$) ranges are equal, while the consequences of unequal scales will be discussed later. Our first observation is the critical temptation level where cooperators dies out is very similar to those values we observed for simulations on lattice-type interaction graphs \cite{szabo_pre05}. There is, however, a conceptual difference from the systems simulated on lattices. As our present plot shows, cooperators become fully dominant as we approach $T=1$ temptation value, which is missing in a lattice-type environment. For example on square lattice cooperators control only the two thirds of the whole population even at $T=1$ because the strict degree number rule of nodes allows defectors to coexist with cooperators \cite{doebeli_el05}. But in our off-lattice environment there is no such artificial constraint and the advantage of spatial reciprocity enjoyed by cooperator strategy manifests entirely. Our plot also warns us that by choosing too large characteristic length scale the system behavior becomes practically identical to the one observed for well-mixed population. In particular, if $l_i=l_b \ge 0.8$ then off-lattice simulations reproduce the mixed, unstructured results. But if we keep the mentioned values low then off-lattice simulations reveal the behavior of structured populations.

\begin{figure}
\centering
\includegraphics[width=7.8cm]{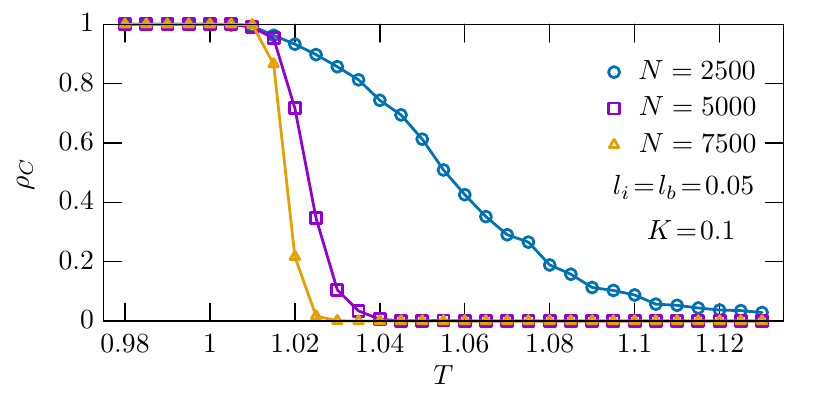}\\
\caption{Cooperation level as a function of temptation value for different density of players. The total number of inhabitants are marked in the legend. Rare population provides better chance for cooperation strategy.}\label{density}
\end{figure}

Our second plot shown in Fig.~\ref{density} summarizes how the total number of inhabitants modify the system behavior. When the population is rare then the cooperation level is significantly higher compared to the cases when we increase the average number of players. For example the critical threshold value of temptation is about $T=1.03$ for $N=7500$, while the same temptation value provides a cooperator dominance for the $N=2500$ case. 
We should emphasize that the improvement of cooperation level is conceptually different from the one observed on lattice simulations where there was an optimal intermediate concentration of players which ensured the highest cooperation level \cite{wang_z_srep12,guan_cp07,wang_z_pre12b}. While the lastly mentioned phenomenon was strongly related to the percolation threshold of the specific lattice structure, it has a different explanation in off-lattice environment.

\begin{figure}[b!]
\centering
\includegraphics[width=7.8cm]{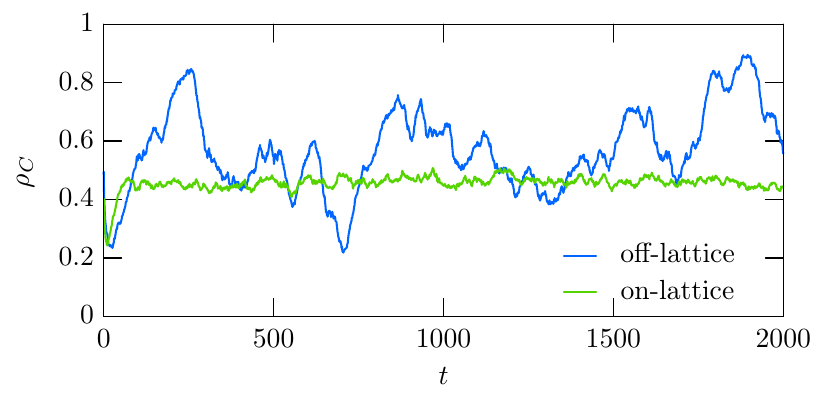}\\
\caption{Time evolution of cooperation level obtained for off-lattice and on square lattice simulations where all other parameters, including temptation value, noise, and total number of inhabitants are equal. In particular, $T=1.015, K=0.1, N=5000$. It is salient that the fluctuations for off-lattice population are significantly larger than in case of on-lattice population.}\label{fluctuations}
\end{figure}

But before discussing its origin, let us compare the time evolutions of cooperation level for an off-lattice and a square-lattice environment. Importantly, all other parameters, including the temptation value, the noise level, the number of players are identical to both cases. Figure~\ref{fluctuations} demonstrates clearly that there is a strong fluctuation in off-lattice environment compared to lattice-based simulations and the system travels between the full cooperator and full defector states permanently. 
The described phenomenon can be monitored in the attached video where the color of population changes almost periodically \cite{5000}. This animation also helps us to understand the origin of this heavy oscillation. Importantly, the borders separating homogeneous domains are not as sharp as for graph-based simulations. This soft intermediate zone allows defectors to crack the phalanx of cooperators which would be robust and steady in a lattice population. As a result, homogeneous blue patches of cooperators diminish eventually giving room for red defector players. A homogeneous red patch, however, becomes vulnerable, too, because in the absence of cooperators defectors cannot exploit their neighbors and they are unable to collect competitive payoff. 

The above mentioned separating zone becomes empty for small $N$ values which has a crucial consequence especially for higher temptation value. Here a homogeneous island may become isolated from the rest of the population, hence resulting in a coexistence of competing strategies. In this frozen state cooperators can survive even if the relatively high temptation value would dictate a full defector state on a lattice.

\begin{figure}
\centering
\includegraphics[width=7.8cm]{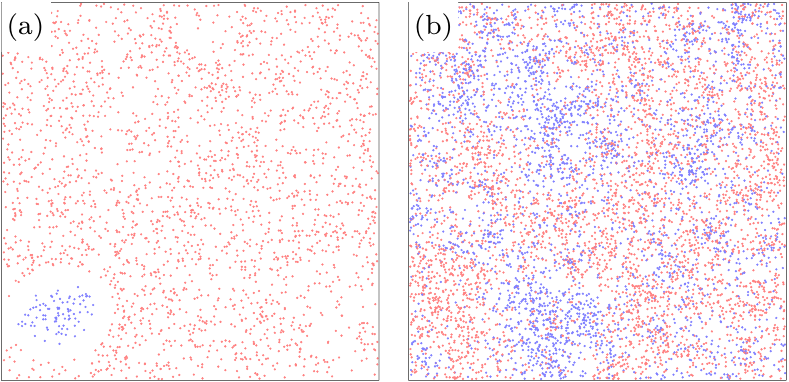}\\
\caption{Spatial distribution of players in a coexistence phase. In both panels defectors are marked by red while cooperators are denoted by blue color. In a rare population, shown in panel~(a), strategies can be separated easily resulting in a frozen final state. This phenomenon makes possible for cooperators to survive even at a relatively high temptation value, as shown in Fig.~\ref{density}. When the average density of population is high, illustrated in panel~(b), then competing strategies maintain a dynamical equilibrium. The latter state is common in on-lattice populations. Parameters are $N=2500, b=1.03$ for panel~(a), and $N=7500, b=1.01$ for panel~(b). In both cases $l_i=l_b=0.05$ were applied.}\label{snapshots}
\end{figure}

The two types of coexistence are illustrated in Fig.~\ref{snapshots}. In panel~(a) we have plotted the above mentioned frozen state which can be observed for small $N$ and high $T$ values. Technically the competing strategy ``coexist", albeit they have no proper interactions. Panel~(b) illustrates a snapshot for a more crowded population at a smaller temptation value where the coexistence of strategies is a  dynamical process similar to those we described regarding the animation.  

\begin{figure}
\centering
\includegraphics[width=7.8cm]{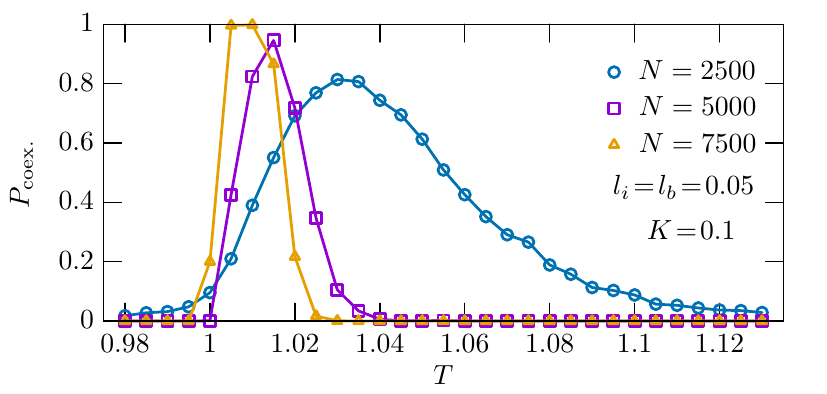}\\
\caption{The chance of coexistence between competing strategies in dependence of the temptation value for different numbers of inhabitants.}\label{coex-PD}
\end{figure}

The above described differences between the pattern formation in rare and crowded population can be made more quantitative if we plot the probability of coexisting state for populations containing different numbers of inhabitants. This is shown in Fig.~\ref{coex-PD}, where we can see that crowded population makes the coexistence harder. Either cooperators, or defectors prevail depending on the temptation value. Less busy population, however, can easily result in isolated patches, hence providing an escape route for cooperators at harsh temptation values.

Next we briefly summarize the possible impact of noise parameter in off-lattice environment. This question could be specially interesting, because earlier observations revealed that some character of the interaction topology could be a decisive factor how the critical threshold value depends on the noise parameter \cite{vukov_pre06}. More precisely, if there are overlapping triangles in the interaction graphs then the critical threshold value where cooperators die out is a decaying function of noise parameter. In the lack of it the mentioned function is non-monotonous and has an optimum at an intermediate noise strength. This optimal level can be the result of a selection process \cite{szolnoki_pre09d}. 

\begin{figure}
\centering
\includegraphics[width=7.8cm]{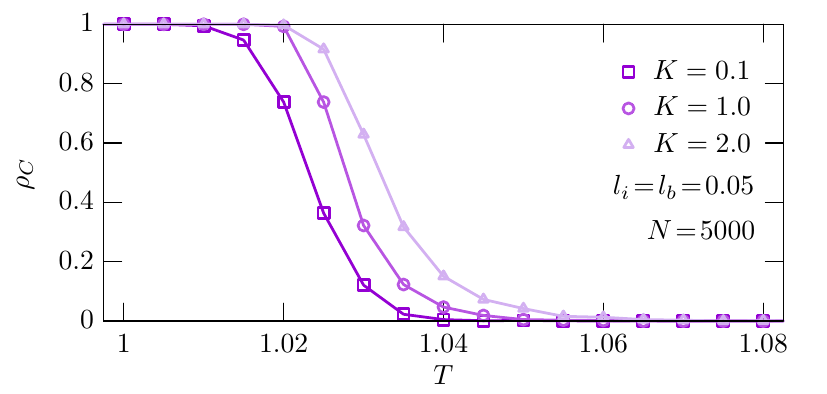}\\
\caption{Noise dependence of cooperation level for prisoner's dilemma game at fixed $N$ and $l$ values. By increasing the uncertainty of imitation the cooperation becomes more likely even at higher temptation values.}\label{noise}
\end{figure}

In on off-lattice simulation, in the absence of a characteristic interaction topology, we observed a different situation which represents a new type of behavior. As Fig.~\ref{noise} demonstrates, in off-lattice environment a higher noise level ensures a better chance for cooperators to survive. This effect is again related to the above described isolation process. At higher noise level the prompt invasion of defectors at larger temptation is not straightforward, which makes a chance for cooperators to survive the first attack. Later they have a higher chance to be isolated, hence to maintain a modest cooperation level.

\begin{figure}[b!]
\centering
\includegraphics[width=7.8cm]{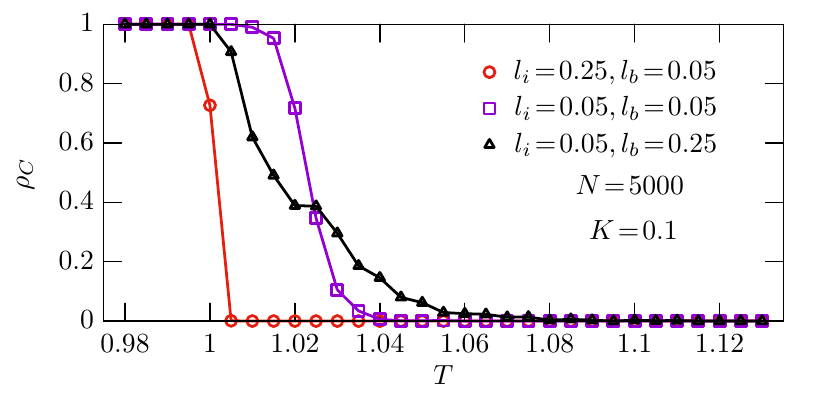}\\
\caption{Cooperation level in dependence of temptation value for smaller and larger range of birth process as indicated in the legend. The chance of remote birth of newcomer weakens the cooperation for moderate $T$, but increase it for higher $T$ values. This phenomenon is robust for other $l_b$ values, too.}\label{large_birth}
\end{figure}

In the following we utilize the liberty of our model and allow the defined length scales to be different. Our key observations are summarized in Fig.~\ref{large_birth}. First, we should note that the comparison is a bit misleading because it suggests that $l_b < l_i$ destroys significantly the cooperation level. But the proper reason of this behavior is the pretty high value of $l_i = 0.25$, which brings the system towards the well-mixed condition. Normally, when $l_i$ is below 0.1 then the application if smaller $l_b$ value does not change the cooperation level significantly. 

However, this is not valid when $l_b$ exceeds the current $l_i$ value. A typical curve is shown in Fig.~\ref{large_birth}, which suggests that the increase of $l_b$ has two-fold consequences. When the temptation is moderate and the cooperator strategy would be dominant then an enhanced $l_b$ helps defectors to jump into the bulk of a cooperator island, which becomes more vulnerable in this way. Consequently, the cooperation level decays comparing to the $l_i=l_b$ case. In contrast, for higher temptation the impact has the opposite sign. Here the system would evolve toward a full defector state in the default case. But a higher $l_b$ might help cooperators to ``escape'' from defectors. When such a remote patch becomes homogeneous then it can survive. In other words, increasing $l_b$ has a similar impact on the system evolution to the behavior we observed for smaller $N$ in the default $l_i=l_b$ case.

\begin{figure}
\centering
\includegraphics[width=7.8cm]{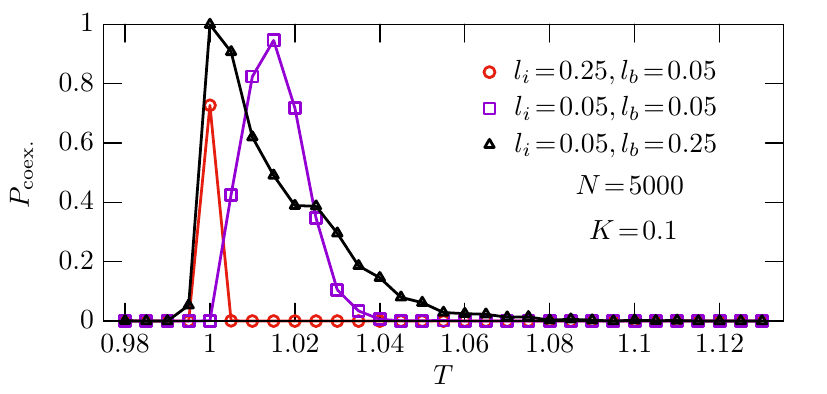}\\
\caption{The probability of coexisting state in dependence of temptation value for normal and enlarged birth range. For moderate $T$ values enhanced birth distance reduces, while for higher temptation values it boosts the chance of a two-strategy final state.}\label{coex_lb}
\end{figure}

The above described argument can be supported nicely if we measure the probability of coexistence state in dependence of temptation. The comparison can be seen in Fig.~\ref{coex_lb} where we can observe similar behavior we presented in Fig.~\ref{coex-PD}. Accordingly, the message is clear: for moderate temptation the enhanced reproduction range prevents forming homogeneous (cooperative) state, while for larger temptation values it helps to maintain coexistence state, hence to increase the average cooperation level.

Summing up our observations, the off-lattice environment provides mostly similar evolutionary trajectories to those previously reported for lattice-based simulations, but the consequence of spatial reciprocity could be stronger and the coexistence of competing strategies is more dynamical in the former case. 

\begin{figure}
\centering
\includegraphics[width=7.8cm]{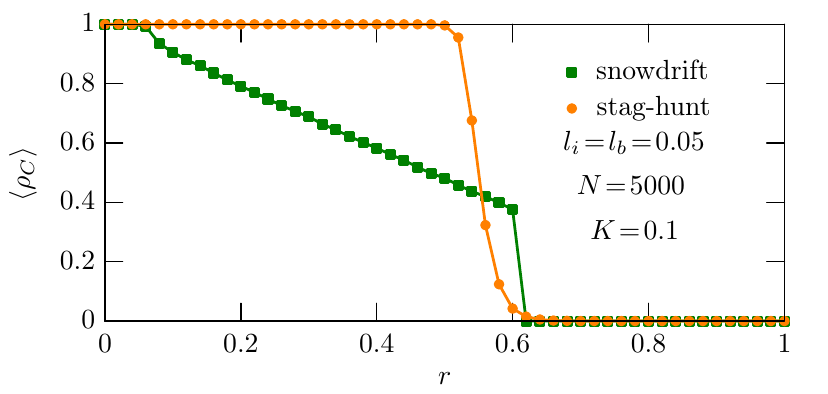}\\
\caption{Cooperation level for snowdrift game and for stag-hunt game where the applied control parameter crosses the related quadrants of $T-S$ diagonally. Similarly to the lattice-based environments the transition from full $C$ to full $D$ state is gradual in the first case and sharp in the second case.}\label{SD-SH}
\end{figure}

These conclusions remain intact when we leave the prisoner's dilemma game and consider snow-drift and stag-hunt games. An illustration can be seen in Fig.~\ref{SD-SH} where we plotted the cooperation levels in dependence of the control parameter $r$ which makes possible to cross the related quadrant of $T-S$ parameter plane diagonally. The green square symbols show a gradual decay of cooperation level for snow-drift game which agrees with previous observations in lattice populations \cite{hauert_n04,li_pp_pre12,shu_f_pa18}. In sharp contrast to this in the stag hunt game the orange circle symbols sign a sharp transition from the full cooperator to the full defector state as we change the control parameter. But this feature is again in good agreement with the reported behavior of on-lattice populations \cite{roca_plr09,starnini_jsm11,szabo_jtb12,wang_l_csf13}.

\section{DISCUSSION}
\label{sec:discussion}

The application of graphs to describe interactions of multi-agent systems becomes an extremely vibrant and successful theory in the last two decades. Not really surprisingly, evolutionary game theory has also enjoyed the benefit of this approach and utilized its concepts and simulation techniques to model more realistic dilemma situations \cite{szabo_pr07,perc_pr17}. Let us stress very clearly that to understand collective behaviors based on graph-based modeling offers not just a simpler technique, but it is proved to be appropriate concept in several real-life systems. But there are cases where off-lattice simulations seem to be more appropriate, hence we can not avoid the numerical difficulties of the latter models. We just quote here some microbiological systems, but other situations, like collective movement or floating may also require off-lattice modeling \cite{vicsek_prl95,avelino_epl18,avelino_epl20,bazeia_epl20}.
 
Our present work illustrates nicely that off-lattice systems where interaction are described by social dilemmas behave conceptually similar to those we observed on lattice-based populations. Therefore the latter, which are numerically more feasible, could be a reliable tool to explore the collective behavior of spatial populations. There are some minor differences, however, which warn us that not all predictions of lattice-based models are robust enough to apply in general. For example, in a system where the crowding is not really limited the so-called spatial reciprocity may work more strongly. We note that the role of aggregation was also reported in another work where the comparison of on- and off-lattice populations were also studied in a different system \cite{daly_nc19}. But staying at our social dilemma systems, the isolation of subgroups could also be a phenomenon which has relevant consequence on the system behavior.

One may claim that we only studied dilemmas which are based on pair-interactions of players and a system with multi-point interactions may behave differently in off-lattice environment. Indeed, a graph-based population ruled by public goods game may behave slightly differently from populations driven by prisoner's dilemma \cite{perc_jrsi13}. But on the other hand, even multi-point interactions were identified as a key factor to diminish the differences of graph topologies \cite{szolnoki_pre09c,szolnoki_pre14c,szolnoki_njp15}. We therefore believe that conclusions obtained off-lattice simulations remain valid for games based on multi-point interactions, but future works can confirm it in more detail.

\section*{ACKNOWLEDGMENTS}

B.F.O. thanks Funda\c c\~ao Arauc\'aria, and INCT-FCx (CNPq/FAPESP) for financial and computational support.

\bibliographystyle{elsarticle-num-names}

\end{document}